\documentclass[12pt,english,floatfix,nofootinbib,superscriptaddress,aps,prd,preprint]{revtex4}
\usepackage[english]{babel}
\usepackage[utf8]{inputenc}
\usepackage{amsfonts}
\usepackage{amsmath}
\usepackage{amssymb}
\usepackage{graphicx,color}
\usepackage[bookmarks=true,colorlinks=true,linkcolor=blue,citecolor=blue,pdfstartview=FitV,urlcolor=blue,breaklinks=true]{hyperref}
\usepackage{indentfirst}
\usepackage{wrapfig}
\usepackage{float}
\usepackage{units}
\usepackage{textcomp}
\usepackage{wasysym}
\usepackage{multirow}
\usepackage{subfig}
\usepackage{array}
\usepackage{epsfig}
\usepackage{cancel}
\usepackage{stmaryrd}
\usepackage{upgreek}
\usepackage{tabularx}
\usepackage{psfrag}
\usepackage{epsfig}
\usepackage{titletoc}
\usepackage{mathrsfs}
\usepackage[dvips]{feynmp}
\usepackage{fancyhdr}
\usepackage[dvipsnames]{xcolor}

\usepackage{hyperref}
\hypersetup{
    colorlinks,
    citecolor=blue,
    filecolor=green,
    linkcolor=purple,
    urlcolor=red,
}

\newcommand{\ie}{\begin{equation}}
\newcommand{\fe}{\end{equation}}
\newcommand{\se}{\begin{eqnarray}}
\newcommand{\ff}{\end{eqnarray}}

\begin{document}

\title{Quantum gases on a torus}  

\author{A. A. Ara\'{u}jo Filho}
\email{dilto@fisica.ufc.br}

\affiliation{Universidade Federal do Cear\'a (UFC), Departamento de F\'isica,\\ Campus do Pici,
Fortaleza -- CE, C.P. 6030, 60455-760 -- Brazil.}
\affiliation{Departamento de Física Teórica and IFIC,
Centro Mixto Universidad de Valencia--CSIC. Universidad
de Valencia, Burjassot-46100, Valencia, Spain}

\author{J. A. A. S. Reis}
\email{jalfieres@gmail.com}

\affiliation{Programa de Pós-graduação em Física, Universidade Federal do Maranh\~{a}o (UFMA),\\ Campus Universit\'{a}rio do Bacanga, S\~{a}o Lu\'{\i}s -- MA, 65080-805, -- Brazil}

\affiliation{Universidade Estadual do Sudoeste da Bahia (UESB), Departamento de Ciências Exatas e Naturais, Campus Juvino Oliveira, Itapetinga -- BA, 45700-00,--Brazil}

\author{Subir Ghosh}
\email{subirghosh20@gmail.com}

\affiliation{Physics and Applied Mathematics Unit, Indian Statistical Institute,\\ 203 B.T.Road, Kolkata 700108, India.}

\date{\today}

\begin{abstract}

This manuscript is aimed at studying the thermodynamic properties of quantum gases confined to a torus. To do that, we consider \textit{noninteracting} gases within the grand canonical ensemble formalism. In this context, fermoins and bosons are taken into account and the calculations are properly provided in both analytical and numerical manners. In particular, the system turns out to be sensitive to the topological parameter under consideration: the winding number. Furthermore, we also derive a model in order to take into account \textit{interacting} quantum gases. To corroborate our results, we implement such a method for two different scenarios: a ring and a torus.

\end{abstract}

%\pacs{PACS Numbers}
%\keywords{Thermodynamic Properties; Non-Cartesian Geometries; Grand Canonical Ensemble; Non-Interacting and Interacting Quantum Gases; Spinless, Bosons and Fermion Particles.}
\maketitle

%%%%%%%%%%%%%%%%%%%%%%%%%%%%%%%%%%%%%%%%%%%%%%%%%%%%%%%%%%%%%%%%%%%%%%%%%%%%%%%%%%%%
\section{Introduction}

The studies concerning the thermodynamic properties of materials have received substantial attention in past years particularly within the context of development of new materials \cite{keshavarz2021properties,i1,paulson2019bayesian,si2021magnetic,i8,i2,i3,leong2019nano}. Based on some well-known approximations, electrons of a given metal may be studied as a gas \cite{lee1988development,araujo2017,aa1,aa2,aa3,silva2018} and such electron systems are worthy to be examined because of their significance in applied \cite{i4,i6} and in fundamental \cite{i5,i7} physics.

The literature possesses a long-term problem in quantum mesoscopic systems, which is the accomplishment of the sum over all states considering both \textit{noninteracting} and \textit{interacting} particle modes. In many cases, the boundary effects cannot be neglected; rather, they must be considered for the sake of obtaining an agreement with experimental results. Naturally, the properties of some physical systems are deemed to be shape dependent \cite{i10,Dai2003,Dai2004} and responsive to their topology \cite{t1,t2,t3,ada2,t5}.

In parallel, regarding a theoretical point of view, there exists an analogous problem in statistical mechanics, which is carrying out the summation over all quantum states in order to acquire physical quantities \cite{pathria1972statistical,landau2013statistical,ada1}. Mainly, the spectra of the particle states are obtained by studying their respective boundary effects. However, when the particle wavelength is too small, if compared to the system under consideration, the boundary effects may be neglected.

In this sense, this manuscript has the purpose of investigating the thermal aspects of quantum gases confined to a torus. To do so, we consider \textit{noninteracting} gases within the grand canonical ensemble approach. Within this context, bosons and fermoins are taken into account, and the calculations are supplied in analytical and numerical ways. Particularly, the system turns out to be responsive to the topological parameter: the winding number. More so, we also derive a model for the sake of taking into account \textit{interacting} quantum gases. Additionally, we implement such a method for two different situations: a ring and a torus

%Initially, in Section \ref{2}, we present a discussion involving the spectral energy for different geometries. Afterwards, in Section \ref{3}, we focus on spinless particles using a setting of the canonical ensemble. Next, in Section \ref{Sec:BF}, we focus on \textit{noninteracting} gases (fermions and bosons) within the same geometries with the usage of the grand canonical ensemble though. Moreover, in Section \ref{6}, we propose two applications towards such a direction: the \textit{Bose-Einstein condensate} and the \textit{helium dimer}. Next, in Section \ref{Sec:Interaction}, we devise a model to perform the {\color{red}calculation} of \textit{interacting} quantum gases which is applied to three different cases: a cubical box, a ring and a torus. These latter three {\color{red}ones} turn out to be more prominent since all the results were derived analytically. Finally, in Section \ref{conclusion}, we conclude and discuss future perspectives.

%%%%%%%%%%%%%%%%%%%%%%%%%%%%%%%%%%%%%%%%%%%%%%%%%%%%%%%%%%%%%%%%%%%%%%%%%%%%%%%%%%%%%%%%%%%%%%%

\section{Noninteracting formalism}\label{Sec:BF}

Even though interactions of molecules and atoms are investigated in many observational experiments, and various aspects can exclusively be realized and presumed by considering interactions \cite{i11,i22,i33,i44}, many prominent characteristics are well delineated by taking into account rather \textit{noninteracting} systems  \cite{chen1972light,e1,e2,e3,e4,e5,e6,e7,e8,e9,e10,e11,e12}.

Examinations of \textit{noninteracting} constituents of matter have several employments, notably in condensed-matter physics \cite{chen1972light,e1,e2,e3,e4,e5,e6,e7,e8,e9,e12,boseeinteincondensate2}, and chemistry \cite{e10,e11}. Particularly, in the \textit{bulk}, it is usually supposed that the calculation of the energy spectrum as well as the \textit{Fermi-Dirac} distribution can be used to examine how its statistics behaves. This approach is acceptable to describe the system of, for instance, a \textit{noninteracting} electron gas. Moreover, such an assumption is entirely justifiable by the following consideration: if the \textit{Fermi energy} is sufficiently large, the kinetic energy (of electrons) close to the Fermi level, will considerably be higher than the potential energy of the so-called \textit{electron-electron} interaction.

\subsection{Thermodynamic approach}

This section is devoted to show the main features of our model under consideration. In other words, we study \textit{noninteracting} particles with different spins (fermions and bosons) using the grand canonical ensemble formalism. It is worth mentioning that for a better comprehension of the reader, these two distinct particles will be studied the in an individual manner. In this sense, the grand canonical partition function for our case is given by%
\begin{equation}
\Xi =\sum_{N=0}^{\infty }\exp \left( \beta \mu N\right) \mathcal{Z}\left[
N_{\Omega }\right] ,  \label{eq:GarndPartition-function}
\end{equation}%
where $\mathcal{Z}\left[ N_{\Omega }\right] $ is the ordinary canonical partition function, $\beta \equiv 1/\kappa_{B}T$, $\kappa_{B}$ is the Boltzmann constant, $T$ is the temperature, and $\Omega$ is a label that represents a generic quantum state. As one can naturally expect, in Eq. (\ref{eq:GarndPartition-function}), there exists the appearance of the occupation number $N_{\Omega }$, and the chemical potential $\mu$. Once fermions and bosons are assumed within our study, then, we have to make a restriction to the occupation number: $N_{\Omega }=\left\{ 0,1\right\} $ for fermionic and $N_{\Omega
}=\left\{ 0,\ldots \infty \right\} $ for bosonic particles. On the other hand, if an arbitrary quantum state is considered instead, the energy relies simply on the occupation number as
\begin{equation*}
E\left\{ N_{\Omega }\right\} =\sum_{\left\{ \Omega \right\} }N_{\Omega
}E_{\Omega }
\end{equation*}%
where%
\begin{equation*}
\sum_{\left\{ \Omega \right\} }N_{\Omega }=N.
\end{equation*}%
With this, the partition function is written as%
\begin{equation}
\mathcal{Z}\left[ N_{\Omega }\right] =\sum_{\left\{ N_{\Omega }\right\}
}\exp \left[ -\beta \sum_{\left\{ \Omega \right\} }N_{\Omega }E_{\Omega }%
\right],
\end{equation}%
leading to%
\begin{equation}
\Xi =\sum_{N=0}^{\infty }\exp \left( \beta \mu N\right) \sum_{\left\{
N_{\Omega }\right\} }\exp \left[ -\beta \sum_{\left\{ \Omega \right\}
}N_{\Omega }E_{\Omega }\right] ,
\end{equation}%
or, in other words, it can properly be rewritten as
\begin{equation}
\Xi =\prod_{\left\{ \Omega \right\} }\left\{ \sum_{\left\{ N_{\Omega
}\right\} }\exp \left[ -\beta N_{\Omega }\left( E_{\Omega }-\mu \right) %
\right] \right\} .
\end{equation}%
Next, we have to perform the sum over all possible occupation numbers. Thereby, we obtain%
\begin{equation}
\Xi =\prod_{\left\{ \Omega \right\} }\left\{ 1+\chi \exp \left[ -\beta
\left( E_{\Omega }-\mu \right) \right] \right\} ^{\chi }.
\end{equation}%
Note that we have added a convenient notation, i.e., $\chi =+1$ for fermions and $\chi =-1$ for bosons. As it is well-known in the literature, the thermodynamical correlation between micro (ensemble theory) and macro world is accomplished by the grand canonical potential:%
\begin{equation}
\Phi =-\frac{1}{\beta }\ln \Xi .
\end{equation}%
Substituting $\Xi$ in above expression and taking into account the following logarithm property
\ie
\log \left[ \prod\limits_{i=1}^{N}\Psi _{i}\right] =\sum\limits_{i=1}^{N}\log \Psi _{i},
\fe
we obtain%
\begin{equation}
\Phi =-\frac{\chi }{\beta }\sum_{\left\{ \Omega \right\} }\ln \left\{ 1+\chi
\exp \left[ -\beta \left( E_{\Omega }-\mu \right) \right] \right\} .
\label{eq:Gand-potential}
\end{equation}%
In this way, the entropy can straightforwardly be given this simple compact form:
\begin{equation}
S=-\left( \frac{\partial \Phi }{\partial T}\right) _{E_{\Omega },\mu }=-k_{B}\sum_{\left\{ \Omega \right\} }%
\left[\chi^{2}\mathcal{N}_{\Omega }\ln \mathcal{N}_{\Omega }-\chi \left( 1-\chi \mathcal{N}%
_{\Omega }\right) \ln \left( 1-\chi \mathcal{N}_{\Omega }\right)\right],
\end{equation}
where%
\begin{equation*}
\mathcal{N}_{\Omega }=\frac{1}{\exp \left[ \beta \left( E_{\Omega }-\mu
\right) \right] +\chi }.
\end{equation*}%
Using the same idea, we can utilize the grand canonical potential to address other thermal quantities, namely, pressure, heat capacity, mean particle number, mean energy.

Therefore, the obtainment of the thermodynamic properties of our system turns out to be a straightforward task. In summary, this occurs because we only have to perform the sum displayed in Eq. $\left( \ref{eq:Gand-potential}%
\right) $. Unfortunately, in several cases, we are no allowed to perform the sum in a close form -- \textit{analytical} results. For our case, it is no different. A numerical analysis must be invoked to overcome such a situation. Thereby, we can obtain the behavior of all quantities considering mainly low temperature regimes. As we shall see, we focus on the investigation of the thermal quantities in a numerical way.

%%%%%%%%%%%%%%%%%%%%%%%%%%%%%%%%%%%%%%%%%%%%%%%%%%%%%%%%%%%%%%%%%%%%%%%%%%%%%%%%%%%%%%%%%%%%%%%%%%%%%%%%%%%%%%%%%%%%%%%%%%%%%%%%%%%%%%%%%%%%%%%%%%%%%%%%%%%%%%%%%%%%%%%%%%%%%%%%%%%%%%%%%%%%%%%%%%%%%%%%%%%%%%%%%%%%%%%%%%%%%%%%%%%%%%%%%%%%%%%%%%%%%%%%%%%%%%%%%%%%%%%%%%%%%%%%%%%%%%%%%%%%%%%%%%%%%%%%%%%%%%%%%%%%%%%%%%%%%%%%%%%%%%%%%%%%%%%%%%%%%%%%%%%%%%%%%%%%%%%%%%%%%%%%%%%%%%%%%%%%%%%%%%%%%%%%%%%%%%%%%%%%%%%%%%%%%%%%%%%%%%%%%%%%%%%%%%%%%%%

\section{Ideal quantum gas on a torus knot}\label{Sec:Torus}

\begin{figure}[tbh]
\centering
\subfloat[Winding number = 1]{
  \includegraphics[width=9cm,height=7cm]{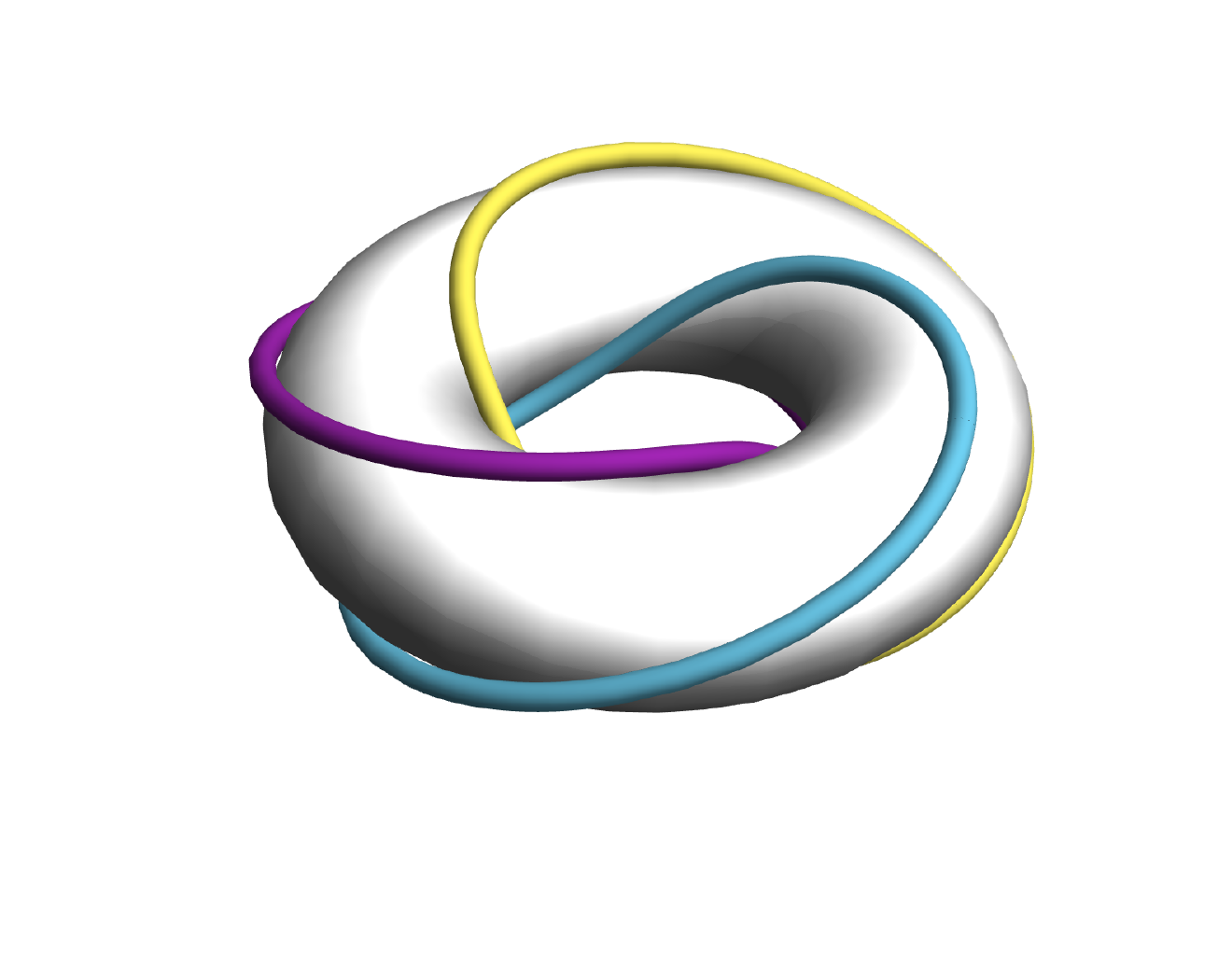}}
\subfloat[Winding number = 2]{
  \includegraphics[width=9cm,height=7cm]{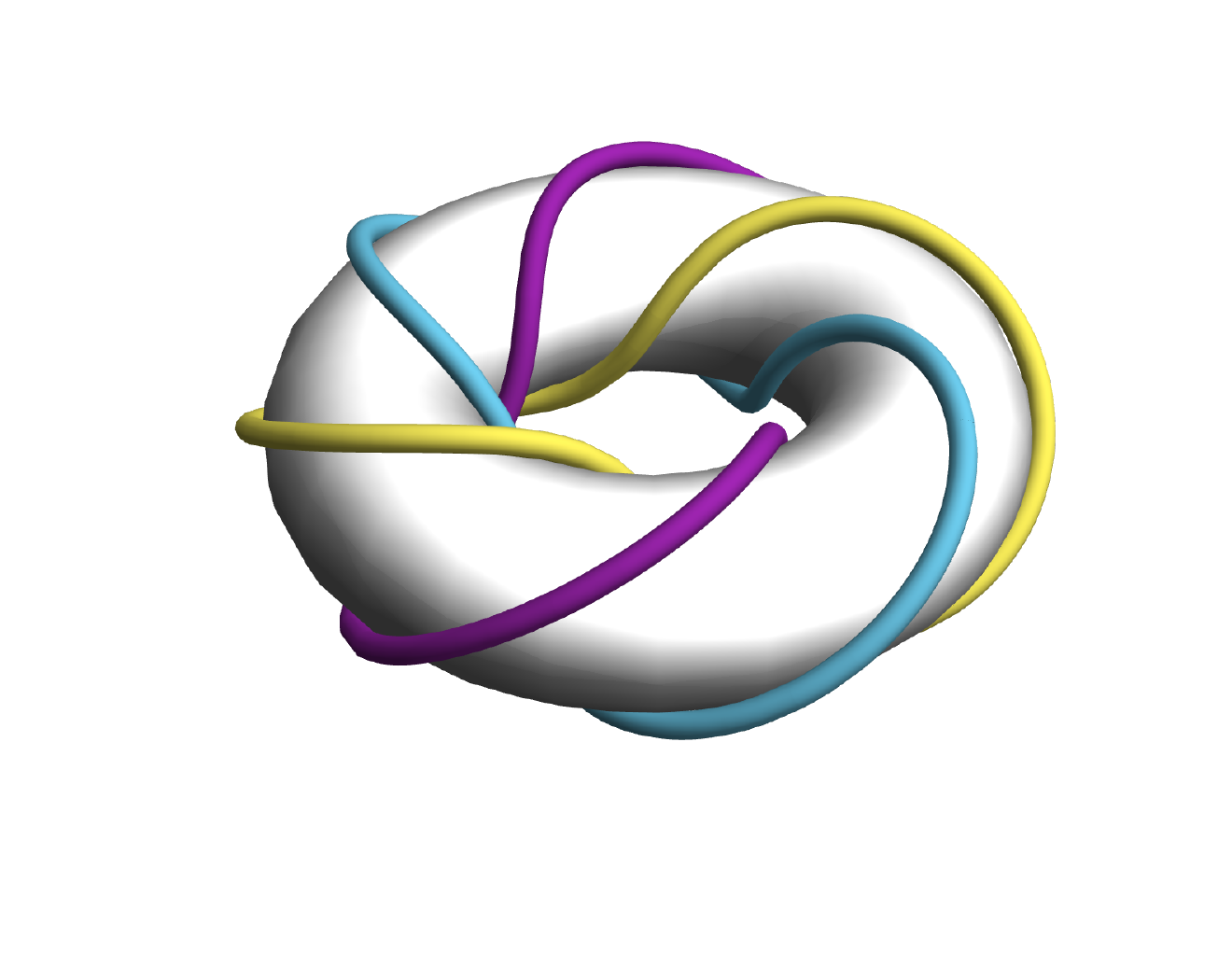}
  }\\
\subfloat[Winding number = 3]{
  \includegraphics[width=9cm,height=7cm]{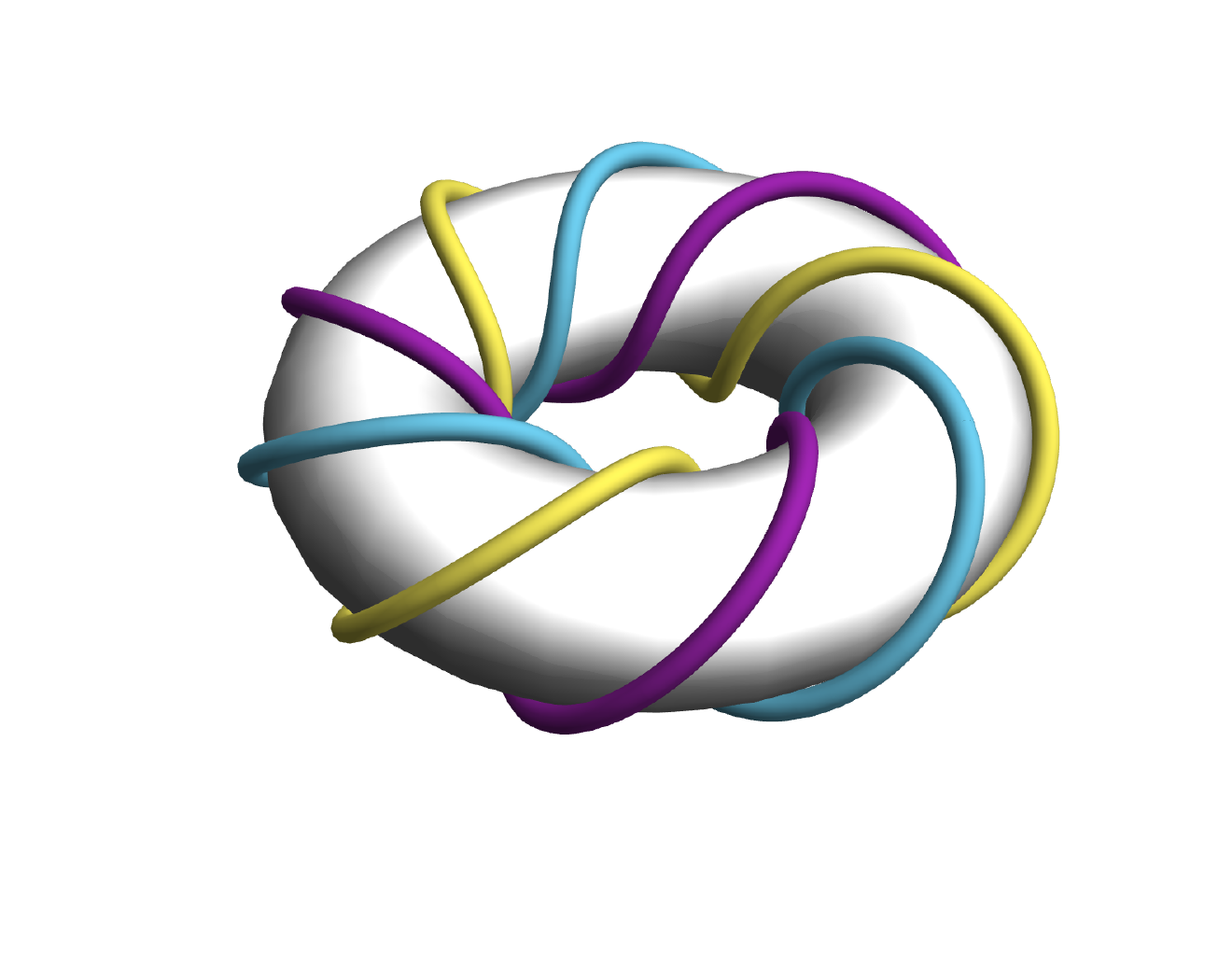}
  }
  \subfloat[Winding number = 4]{
  \includegraphics[width=9cm,height=7cm]{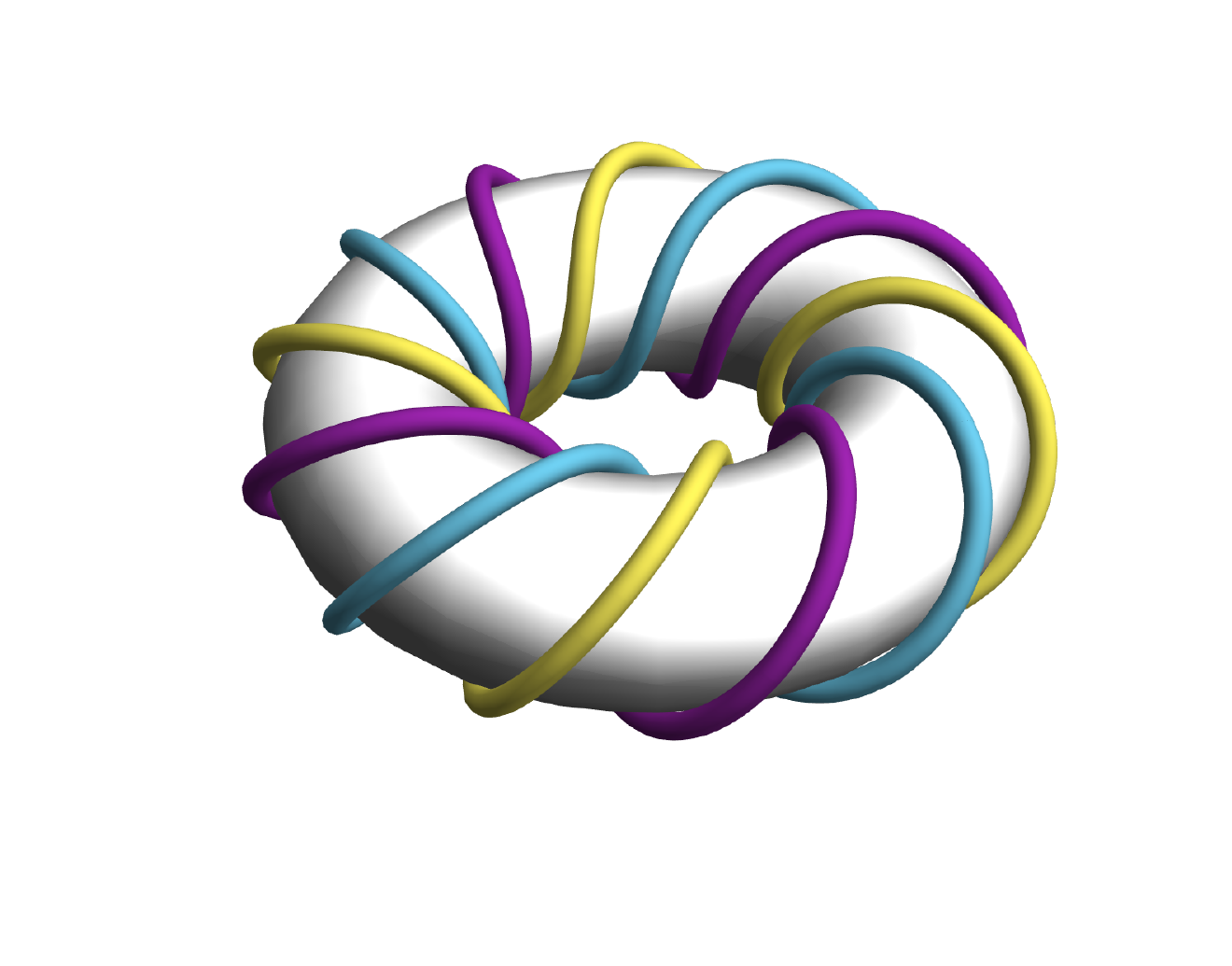}
  }
\caption{Path behavior through a torus knot for different winding numbers }
\label{windingnumbers}
\end{figure}

In past years, the analysis of quantum mechanics within the context of confined systems has received special attention due to its variety of applications: from theoretical through experimental viewpoints
\cite{ttt1,shape1,shape2,shape3,shape4,shape5,shape6,Torus,shape7,shape8}. In particular, it was Freyd et al. who pioneered the study of knot invariants to derive the correlation between pure mathematics and the physical world \cite{jones}. With this idea, a new branch of theoretical(mathematical) physics gave rise to: the topological quantum field theory made by Witten \cite{ttt2,ttt3}. It is important to mention that the relation  between statistical mechanics and invariant knots was also feasable \cite{ttt4}. 
Besides that, there exists another inspiration to explore the physical impact of the torus topology ($\mathrm{T}^{2}= S^{1} \times S^{1}$): the application to living beings focused on the behavior their DNA's \cite{DNA1,DNA2,DNA4,DNA5}. The fundamental group of the torus is given by
\ie
\pi_{1}(\mathrm{T}^{2})= \pi_{1}(S^{1}) \times \pi_{1}(S^{1}) \cong  \mathbb{Z}^{1} \times \mathbb{Z}^{1},
\fe
and furthermore its first homology group is isomorphic to the fundamental group \cite{munkres2018}. On the other hand, in Ref. \cite{Torus}, the respective spectral energy for the torus knot was first calculated. Based on it, we intend to examine how particles of different spins behave in this configuration. In order to obtain this information, we need to solve the Schr\"{o}dinger equation considering the torus potential \cite{Torus}. After that, we acquire the following expression:
\ie
E_{n}^{\text{\textrm{Torus}}} = \frac{n^{2}}{2\mathrm{M} \mathfrak{a}^{2}p^{2}} \frac{\cosh^{2}\eta}{\alpha^{2}+ \sinh^{2}\eta -1},
\label{toruss}
\fe
where $\eta=\cosh^{-1}{R/d}$ is a parameter which fixes the toroidal surface, $\alpha$ is the winding number, $\mathfrak{a}=\sqrt{R^{2}-d^{2}}$, $p$ describes the number of loops in the toroidal direction and $n=0,1,\ldots $. Using these spectra energies, a concise analysis involving the thermal aspects of our system may properly be accomplished in what follows.

For the sake of providing all thermodynamics properties, we have to carried out the sum over the states presented in Eq. $\left( \ref{eq:Gand-potential}\right) $. Favorably, it can be done in a precise manner by the advantage of the well-known \textit{Euler-MacLaurin } formula. Then, the grand partition function to this case is better written as
\begin{equation}
\Phi =-\frac{2\chi}{\beta }\int_{0}^{\infty} \,\ln \left\{ 1+\chi z\exp \left[ -\beta
E\left( n\right) \right] \right\}   \mathrm{d}n+\dots.
\end{equation}%
which follows%
\begin{equation}
\Phi =\frac{ p}{\sqrt{F\left( \alpha ,\eta \right)} }\frac{\mathcal{L}}{\lambda }h_{\frac{3}{2}}\left(z\right) -h_{1}\left(z\right) .  \label{eq:GCanonicalFermion}
\end{equation}%
where $\mathcal{L}=2\pi\mathfrak{a}$, $F(\alpha,\eta) = \frac{\cosh^{2}\eta}{\alpha^{2}+ \sinh^{2}\eta -1}$, $\lambda=\sqrt{\frac{2\pi}{mT}}$, $z=e^{\beta\mu}$ and $h_{\sigma }\left(z\right) $ is defined as
\begin{equation}
h_{\sigma }\left(z\right) =\frac{1}{\Gamma \left(
\sigma \right) }\int_{0}^{\infty }\frac{t^{\sigma -1}}{z^{-1}e^{t}+\chi}\mathrm{d}t.
\end{equation}%
Note that if $\chi=1$, we have the Fermi-Dirac statistics. On the other hand, if $\chi=-1$, we have Bose-Einstein distribution instead. The internal energy reads
\begin{equation}
\mathcal{U}=\chi \frac{\mu z}{T}\log \left( 1+\chi e^{z}\right) +\frac{\lambda \mu z}{T%
}\frac{p\mathcal{L}}{\sqrt{F\left( \alpha ,\eta \right)} }h_{\frac{1%
}{2}}\left( z\right) +\frac{3\lambda }{2}\frac{%
p\mathcal{L}}{\sqrt{F\left( \alpha ,\eta \right)} }h_{\frac{3}{2}%
}\left( z\right)-h_{1}\left( z\right).
\end{equation}
Analogously, the number of particles is written as
\begin{equation}
N=\frac{\chi z}{T}\frac{p\mathcal{L}}{\sqrt{F\left( \alpha ,\eta \right)} }\log
\left( 1+\chi e^{z}\right) +\frac{\lambda z}{T}\frac{p\mathcal{L}}{\sqrt{F\left(
\alpha ,\eta \right)} }h_{\frac{3}{2}}\left( z\right).
\label{eq:particleUI}
\end{equation}%
Moreover, the other thermodynamic state quantities can also be calculated. However, the expressions are lengthy and they are not easy to interpret. Once we shall examine how the thermodynamic quantities are altered when different winding numbers are taken into account, we display Fig \ref{windingnumbers} to show the modification of path through a torus knot. It is important to mention that, from now on, we use $\mu=1~\mathrm{eV}$ for the chemical potential in the numerical evaluation.

\begin{figure}[tbh]
\centering
\subfloat[Energy]{
  \includegraphics[width=8cm,height=5cm]{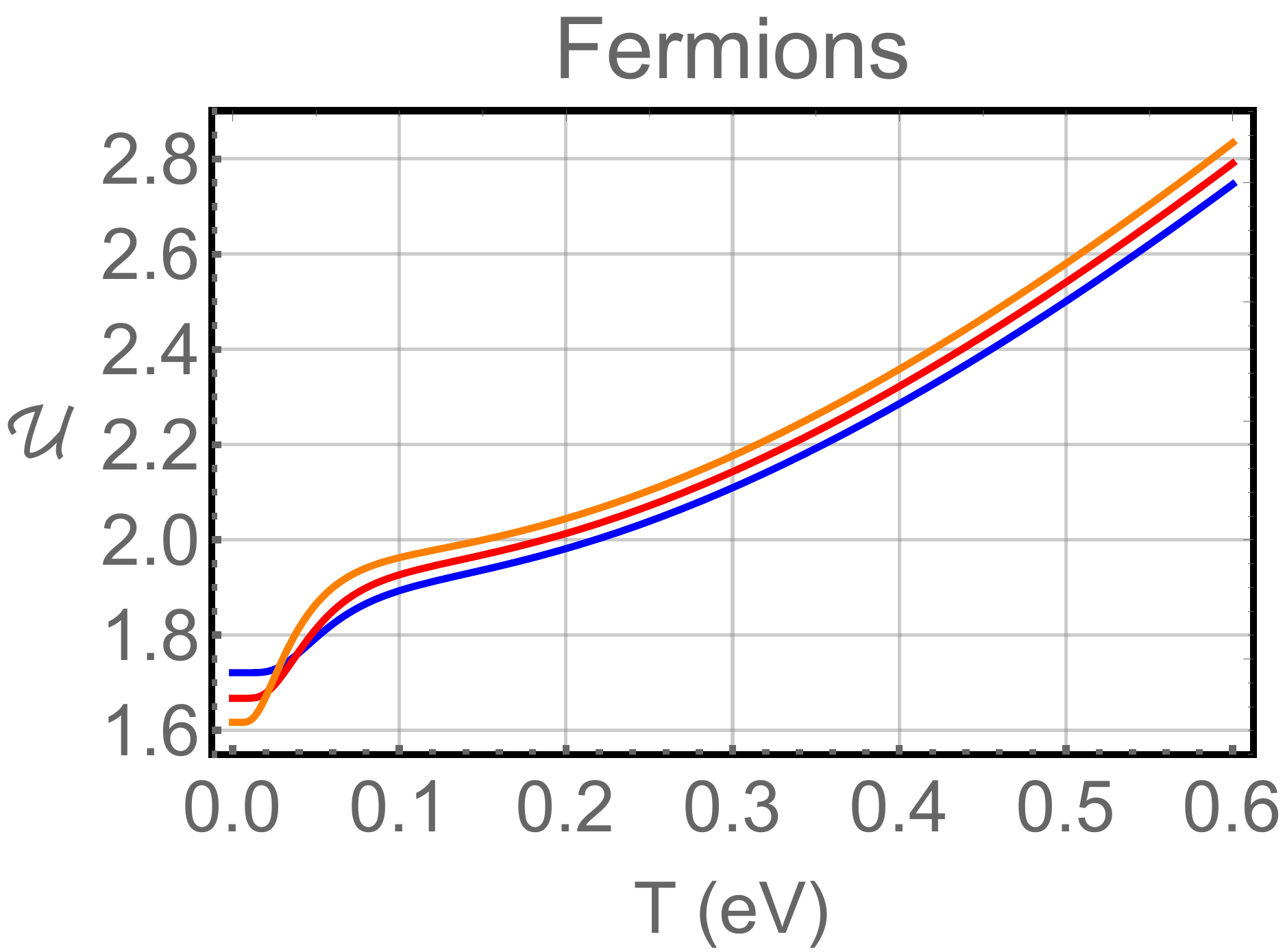}
  \label{fig:T-EC}}
\subfloat[Energy]{
  \includegraphics[width=8cm,height=5cm]{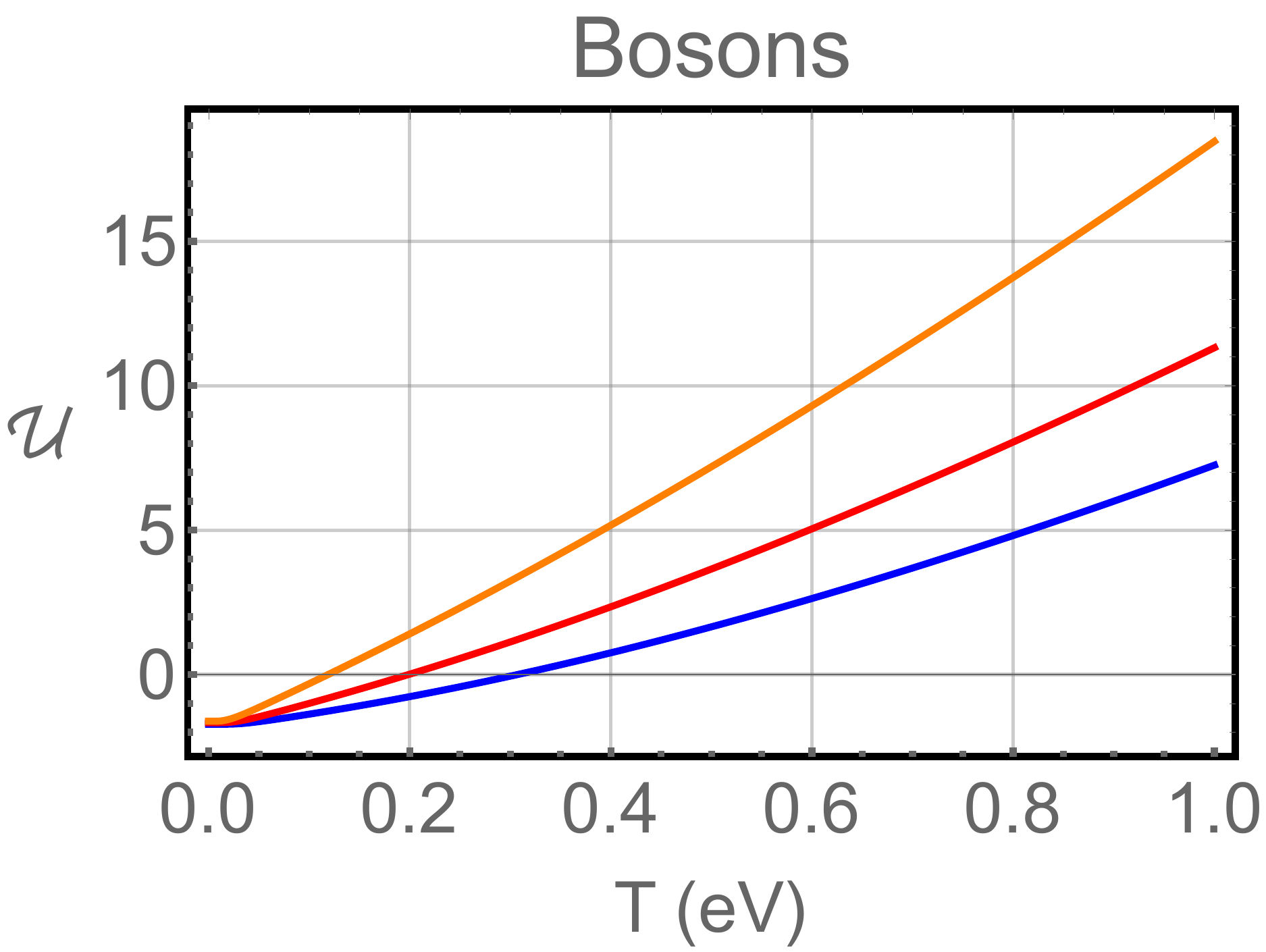}
  \label{fig:T-EE}}
\caption{Energy behavior in the low-temperature regime for different values of winding number $\alpha$ for the torus knot. The blue lines represents $\alpha=1$, the red lines $\alpha=5$ and the orange lines $\alpha=7$.}
\label{fig:Torus-E-LowT}
\end{figure}

At the beginning, the behavior of the mean energy is shown in Fig. \ref{fig:Torus-E-LowT}. Here, we note that fermions and bosons have a substantial difference in their behavior. In one hand, the mean energy of bosons has an increment in its value due to the temperature and the winding numbers. On the other hand, fermions show an “inversion points”. After and before this crucial point, there are distinct behavior for different winding numbers, e.g., below $0.02~\mathrm{eV}$ (the first inversion point), the energy become larger for small winding numbers. As the temperature increases and reach the first inversion point, the energy for $\alpha=7$ becomes equal to the energy for $\alpha=5$. At the second inversion point, at $0.04~\mathrm{eV}$, the energy for $\alpha=1$ becomes equal to the energy for $\alpha=5$. After this point, we realise that the energy for a fixed temperature increases with the topological parameter. This behavior does not occur for bosons. For increasing values of temperature, the energy has to increase as well. Concerning the winding number, the great is the winding number the great is the energy --  which quite different for fermions. One possible explanation to the different behavior between bosons and fermions relies on the Pauli exclusion principle. For instance, $\alpha=1$ corresponds to one single loop to allocate $N$ electrons, whereas for $\alpha=2,3,\ldots$, we have more loops to dish out the same amount of electrons. Then, we expect that the energy becomes greater for small $\alpha's$, since the electrons are confined in a small space. For bosons, on the other hand, we do not have this limitation because Pauli exclusion principle plays no role in this case, i.e., bosons can be condensed in the same state.

In both cases, fermions and bosons, the energy spectrum is the same, then it seems that this inversion point is purely related to the statistic itself. Also, this inversion point is "visible" because we can control the winding numbers. We must remember that it is considered as a free parameter. On the other hand, this analysis is not applied to "angular constraints" where there is no topological parameter that we could control for leading to modifications of the energy values. 

\begin{figure}[tbh]
\centering
\subfloat[Heat capacity]{
  \includegraphics[width=8cm,height=5cm]{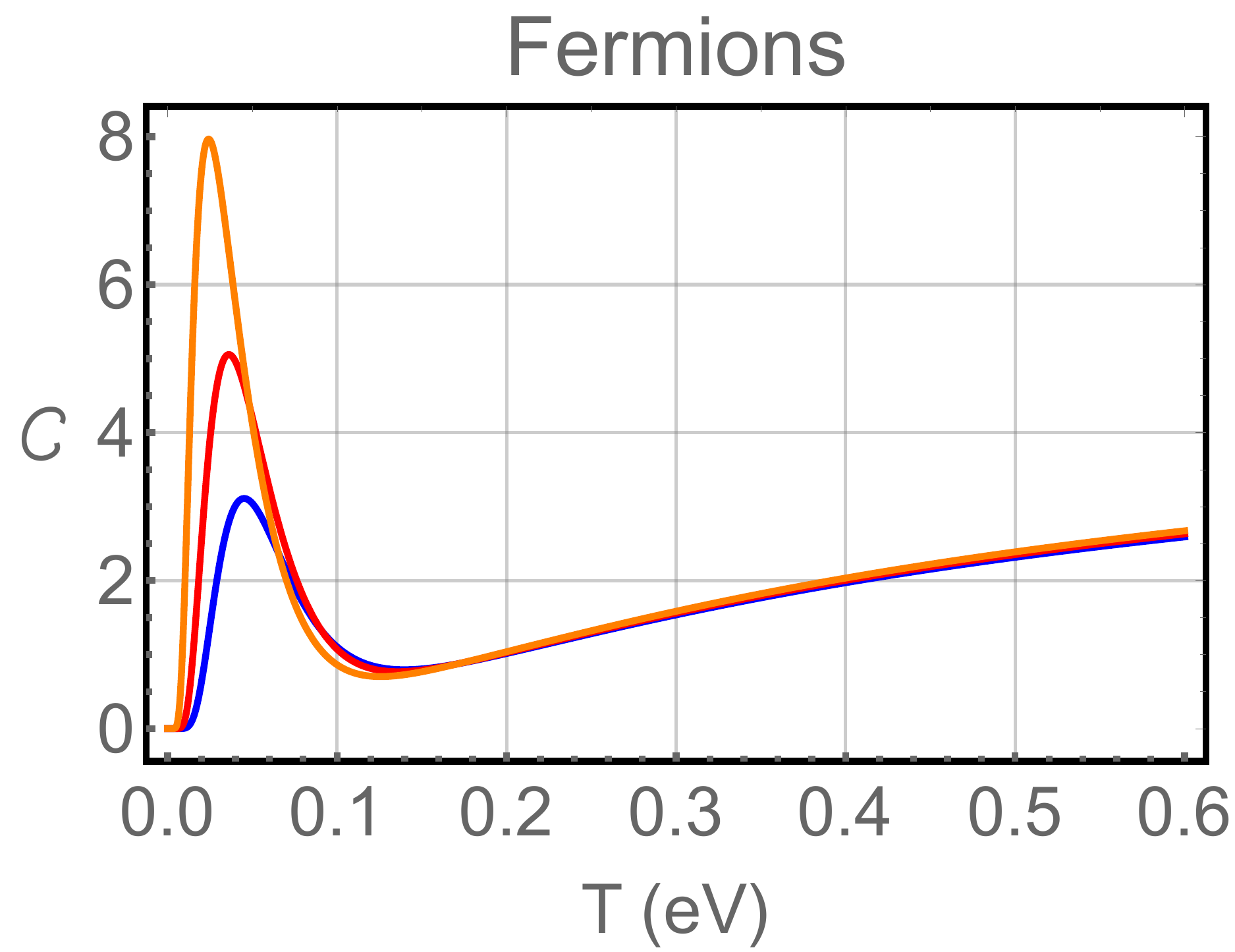}
  \label{fig:T-CC}}
\subfloat[Heat capacity]{
  \includegraphics[width=8cm,height=5cm]{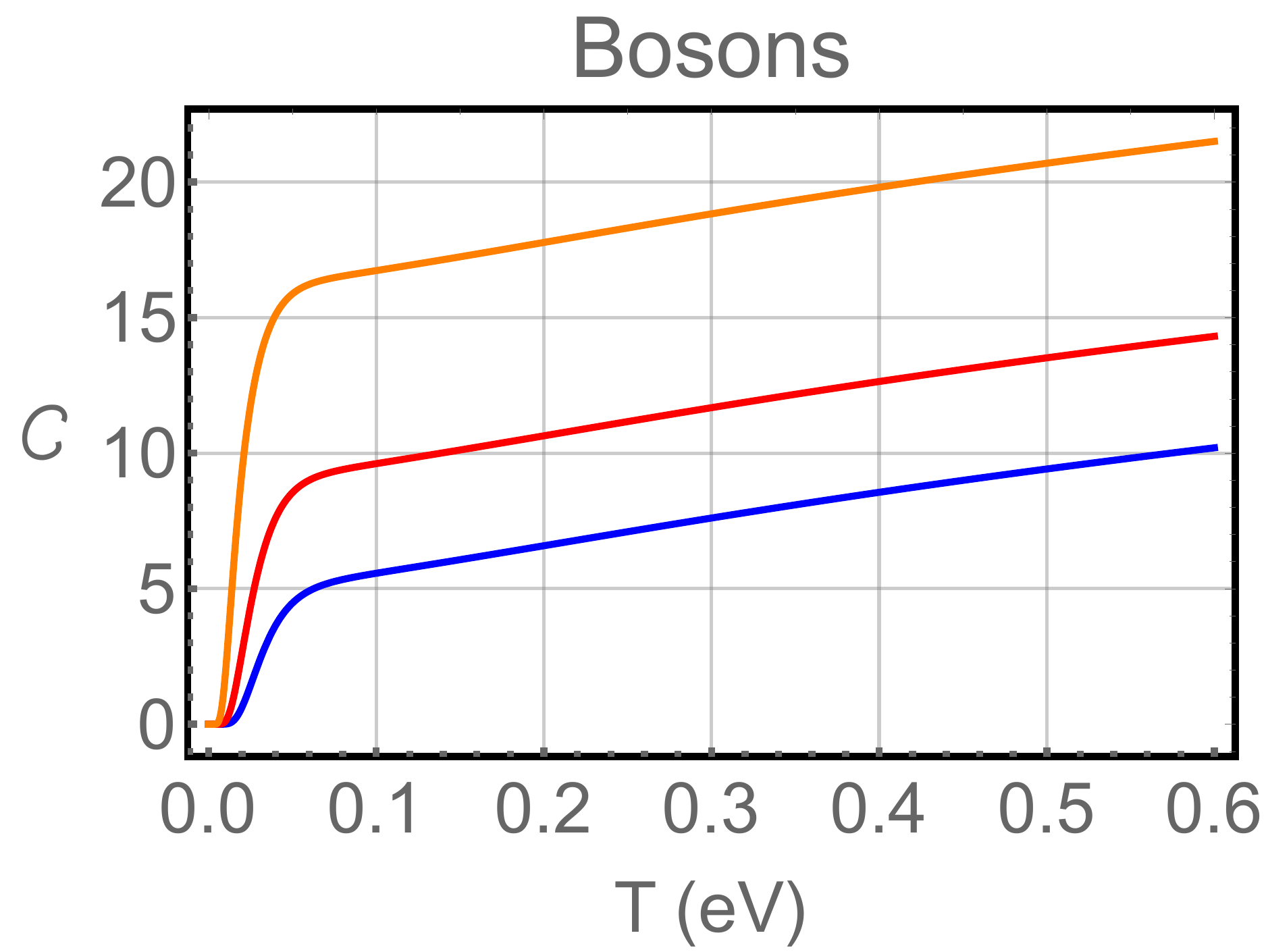}
  \label{fig:T-CE}}
\caption{Heat capacity in the low-temperature regime for different values of winding number $\alpha$ for the torus knot. The blue lines represents $\alpha=1$, the red lines $\alpha=5$ and the orange lines $\alpha=7$.}
\label{fig:Torus-C-LowTx}
\end{figure}

In Fig. \ref{fig:Torus-C-LowTx} , we display the heat capacity considering different spins and winding numbers. For bosons, it is seen that when temperatures are below than $0.05~\mathrm{eV}$, the magnitude of the heat capacity goes faster to zero. On the other hand, when the temperature is above $0.05~\mathrm{eV}$, its magnitude increases. In other words, the great is the winding number the great is the energy. Particularly, for large values of the temperature, the heat capacity tends to a constant value, which also depends on the topological parameter. Lastly, it is observed a notable behavior for fermions. It is noticed that, for temperatures below $1.0~\mathrm{eV}$, the heat capacity possesses a mound that is more expressive when there exists an increment of winding numbers. It is important to mention that above $1.0~\mathrm{eV}$, the heat capacity has its values incremented with temperature until reaching a constant value, which do not depend on the winding number. It is interesting to see a system whose thermodynamic properties depend on a topological parameter. Clearly, this knowledge might be fruitful for further applications in particular system within condensed matter physics.

\begin{figure}[tbh]
\centering
\subfloat[The mean number of particle for fermions]{
  \includegraphics[width=8cm,height=5cm]{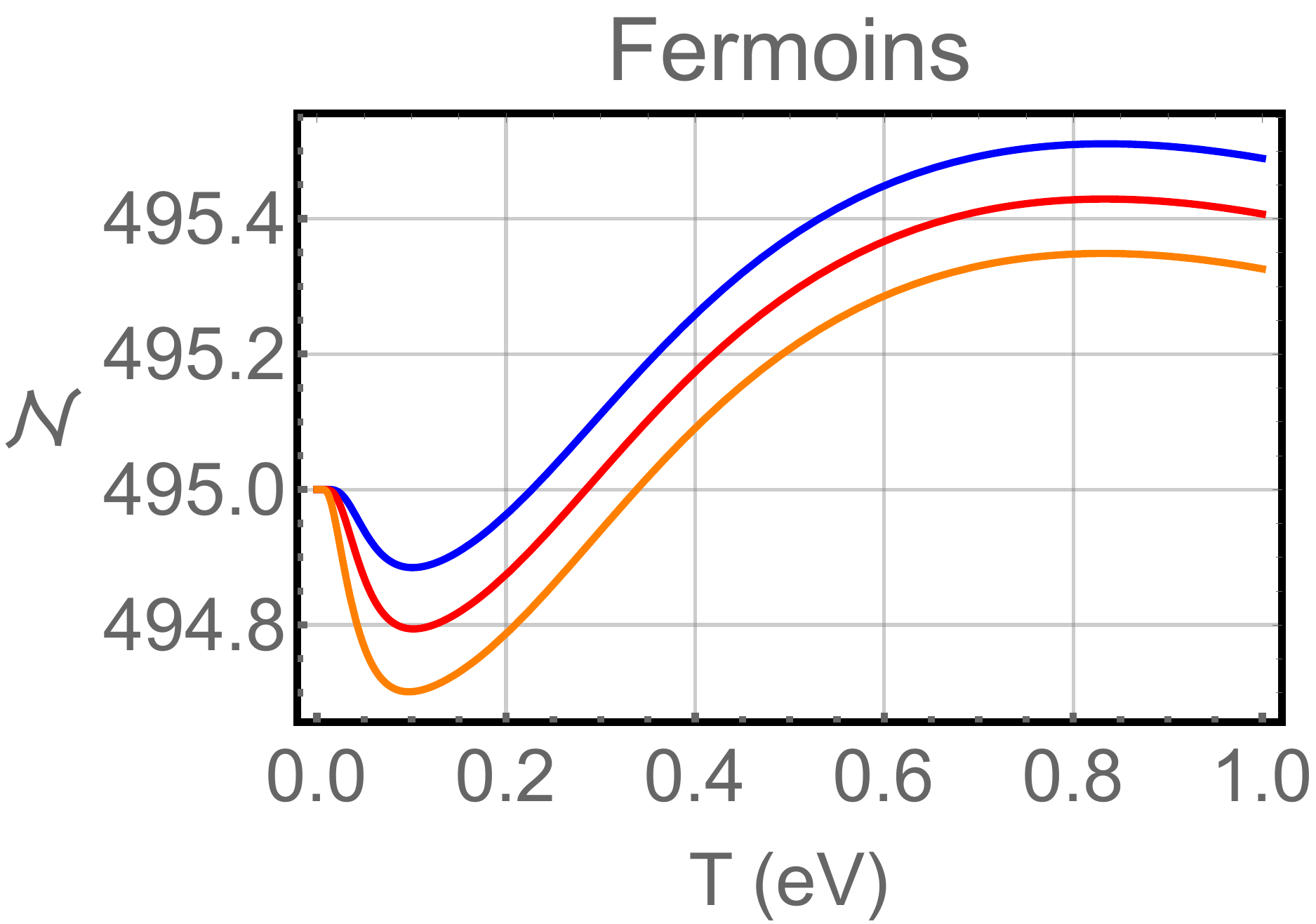}
  \label{fig:NnApp-1}}
\subfloat[The mean number of particle for bosons]{
  \includegraphics[width=8cm,height=5cm]{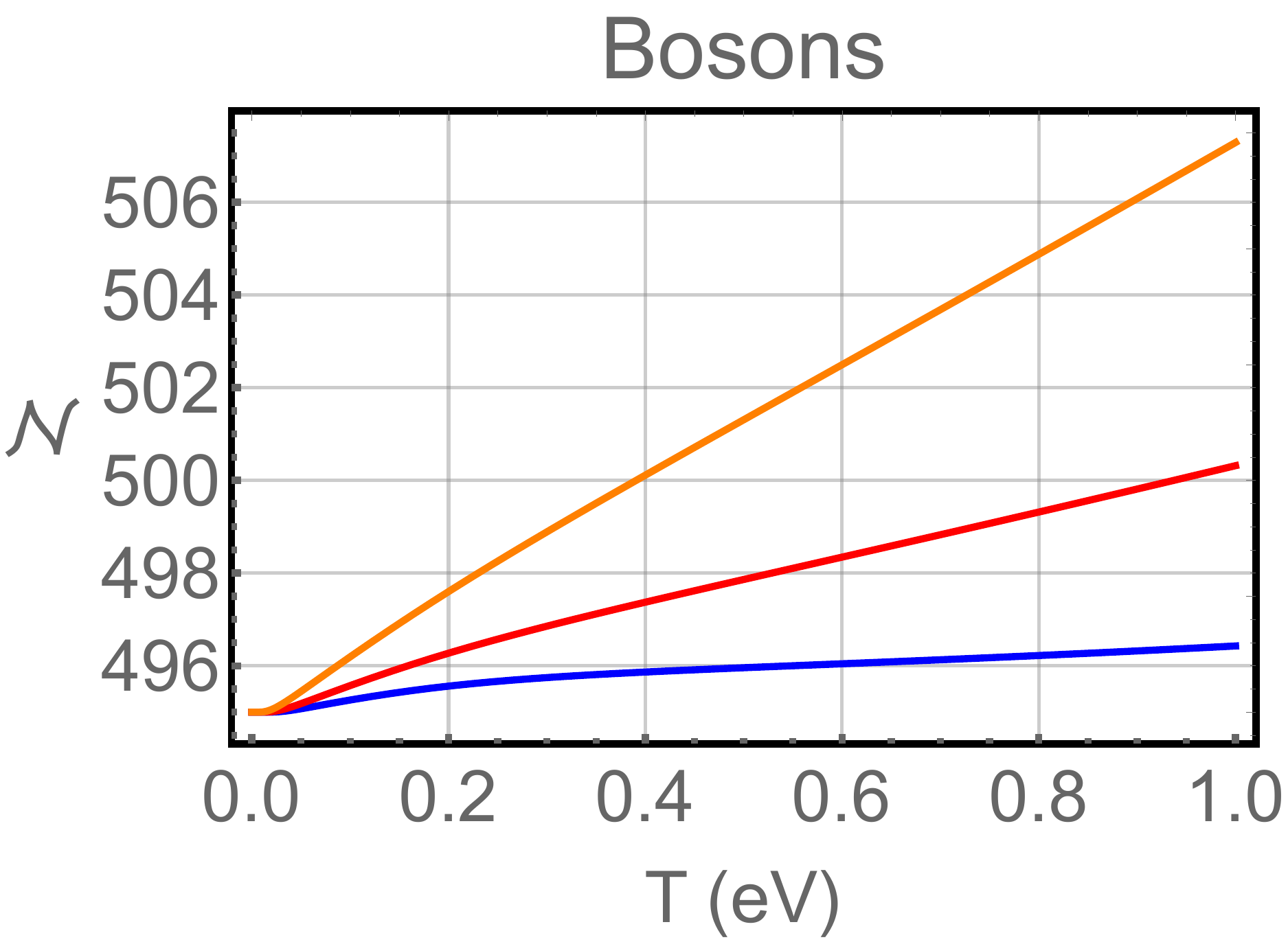}
  \label{fig:NnApp-2}}
\caption{The mean particle number for both fermions and bosons. The thick lines represents the system without interaction where we find blue lines represents $\alpha=1$, the red lines $\alpha=5$ and the orange lines $\alpha=7$. }
\label{fig:NnApp}
\end{figure}

We also analyse how the mean particle number fluctuate for both fermions and bosons, as seen in Fig. \ref{fig:NnApp}. We observe that, in low temperature regime, fermions fluctuations are not too sensitive to temperature variations if compared to bosons. We realize that fermions and bosons have an opposite behavior when there exist modifications in the winding number. For instance, the fluctuations for fermionic particles decrease when we increase the winding number. Bosons, on the other hand, have a growth of fluctuations when we increase the winding number. Those differences are connected with the intrinsic statistics that each particle belongs to. When we look at Eq. \ref{eq:particleUI}, we see that the mean particle number $N$ for bosons and fermions differ in their structure only because of parameter $\chi$. In other words, it controls if we are dealing with either bosons or fermions.

%%%%%%%%%%%%%%%%%%%%%%%%%%%%%%%%%%%%%%%%%%%%%%%%%%%%%%%%%%%%%%%%%%%%%%%%%%%%%%%%%%%%%%%%%%%%%%%%%%%%%%%%%%%%%%%%%%%%%%%%%%%%%%%%%%%%%%%%%%%%%%%%%%%%%%%%%%%%%%%%%%%%%%%%%%%%%%%%%%%%%%%%%%%%%%%%%%%%%%%%%%%%%%%%%%%%%%%%%%%%%%%%%%%%%%%%%%%%%%%%%%%%%%%%%%%%%%%%%%%%%%%%%%%%%%%%%%%%%%%%%%%%%%%%%%%%%%%%%%%%%%%%%%%%%%%%%%%%%%%%%%%%%%%%%%%%%%%%%%%%%%%%%%%%%%%%%%%%%%%%%%%%%%%%%%%%%%%%%%%%%%%%%%%%%%%%%%%%%%%%%%%%%%%%%%%%%%%%%%%%%%%%%%%%%%%%%%%%%%%%%%%%%%%%%%%%%%%%%%%%%%%%%%%%%%%%%%%%%%%%%%%%%%%%%%%%%%%%%%%%%%%%%%%%%%%%%%%%%%%%%%%%%%%%%%%%%%%%

\section{Interacting approach}\label{Sec:Interaction}

\subsection{The model}

This section is to study interactions between the particles under consideration. In this way, we modify briefly the methodology developed in Sec. \ref{Sec:BF}
by adding an interaction term $U\left( V,n\right)$. Notice that for our purpose, this term depends only on the volume $V$ and the particle density $n$. As we shall see, such interaction can be obtained by the well-known mean field approximation. With it, analytical results can be generated. It is noteworthy to be observed that the interaction term depends on the particle density in a monotonic manner. Across this section, the natural units $k_{B}=1$ are adopted. To this case of interactions, the new grand canonical partition function is given by%
\begin{subequations}
\begin{equation}
\mathcal{Z}\left( T,V,\mu \right) =\sum_{\left\{ N_{\Omega }\right\}
=0}^{\left\{ \infty /1\right\} }\exp \left\{ -\beta \left[ \sum_{\left\{
\Omega \right\} }N_{\Omega }\left( E_{\Omega }-\mu \right) +U\left(
V,n\right) \right] \right\} ,  \label{eq:g-partition-function}
\end{equation}%
where
\begin{equation}
z^{N}=\exp \left\{ N\beta \mu \right\} =\exp \left\{ \beta \sum_{\left\{
\Omega \right\} }N_{\Omega }\mu \right\} .
\end{equation}
\end{subequations}%
The summation index appearing
in Eq. (\ref{eq:g-partition-function}), i.e., $\left\{ \infty /1\right\} $, evidences that many bosons can infinitely populate a particular quantum state $\Omega$. On the contrary, for spin-half particles, only one fermion will be allowed because of the Pauli exclusion principle. Compactly, from Eq. (\ref{eq:g-partition-function}), we use the upper index $\infty$ for bosons and $1$ for fermions. Supposing that the interaction term takes the following form: $U\left( V,n\right) =Vu\left( n\right) $. After that, we can write%
\begin{equation}
\mathcal{Z}\left( T,V,\mu \right) =\sum_{\left\{ N_{\Omega }\right\}
=0}^{\left\{ \infty /1\right\} }\exp \left\{ -\beta \left[ \sum_{\left\{
\Omega \right\} }N_{\Omega }\left( E_{\Omega }-\mu \right) +Vu\left(
n\right) \right] \right\} .  \label{eq:g-partition-function-1}
\end{equation}

We assume that $Vu\left( n\right) $ is linear in $\sum_{\Omega
}N_{\Omega }=N$ in order to facilitate the calculations. We perform the Taylor expansion of $u\left( n\right)$ (around the mean value $\bar{n}$) as follows:%
\begin{equation}
u\left( n\right) =u\left( \bar{n}\right) +u^{\prime }\left( \bar{n}\right)
\left( n-\bar{n}\right) +\ldots. \label{eq:Taylor_u}
\end{equation}%
More so, one important fact to be noted is the following: if the potential energy depends only on the position, the \textit{molecular field approximation} will be suitable to be applied from Eq. (\ref{eq:Taylor_u}). This approach is commonly used in the literature specially in the context of condensed matter physics \cite{humphries1972,klein1969,das2016particle,wojtowicz,ter1962molecular,araujo2017,silva2018,tt5,araujo2022does}. The mean energy of the quantum state $\Omega$ is%
\begin{equation}
E=\sum_{\left\{ \Omega \right\} }N_{\Omega }\left[E_{\Omega }+u^{\prime }\left( \bar{n}\right)\right]+U\left( V,\bar{n}%
\right) -u^{\prime }\left( \bar{n}\right) \bar{N}, \label{eq:Total_energy}
\end{equation}%
which entails%
\begin{eqnarray}
\mathcal{Z}\left( T,V,\mu \right) &=&\exp \left\{ -\beta \left[ U\left( V,%
\bar{n}\right) -u^{\prime }\left( \bar{n}\right) \bar{N}\right] \right\}
\notag \\
&&\times \prod_{\Omega =1}^{\infty }\left( \sum_{N_{\Omega }=0}^{\left\{
\infty /1\right\} }\exp \left\{ -\beta \left[ E_{\Omega }+u^{\prime }\left(
\bar{n}\right) -\mu \right] N_{\Omega }\right\} \right) .
\label{eq:Modified_GCP}
\end{eqnarray}%
Differently, we can display above equation as being%
\begin{eqnarray}
\mathcal{Z}\left( T,V,\mu \right) &=&\exp \left\{ -\beta \left[ U\left( V,%
\bar{n}\right) -u^{\prime }\left( \bar{n}\right) \bar{N}\right] \right\}
\notag \\
&&\times \prod_{\Omega =1}^{\infty }\left( 1+\chi \exp \left[ -\beta \left(
E_{\Omega }+u^{\prime }\left( \bar{n}\right) -\mu \right) \right] \right)
^{\chi },
\end{eqnarray}%
where we have considered $\chi=-1$ for bosons, and $\chi=1$ for fermions.

\subsection{Thermodynamic state quantities}

Here, we derive the grand canonical potential in a straightforward manner%
\begin{eqnarray}
\Phi &=&-T\ln \mathcal{Z}  \notag \\
&=&-T\chi \sum_{\Omega }\ln \left( 1+\chi \exp \left[ -\beta \left(
E_{\Omega }+u^{\prime }\left( \bar{n}\right) -\mu \right) \right] \right)
+U\left( V,\bar{n}\right) -u^{\prime }\left( \bar{n}\right) \bar{N}.
\label{eq:grand-canonical-potencial}
\end{eqnarray}
From it, the thermodynamic properties can be acquired. Next, the mean particle number is written
\begin{equation}
\bar{N}=\sum_{\Omega }\frac{1}{\exp \left[ \beta \left( E_{\Omega
}+u^{\prime }\left( \bar{n}\right) -\mu \right) \right] +\chi },
\label{eq:Mean-number-N}
\end{equation}%
and the mean occupation number is $\bar{N}=\sum_{\Omega }\bar{n}_{\Omega }$, where%
\begin{equation}
\bar{n}_{\Omega }=\frac{1}{\exp \left[ \beta \left( E_{\Omega }+u^{\prime
}\left( \bar{n}\right) -\mu \right) \right] +\chi }.
\label{eq:Mean-number-N-new-n1}
\end{equation}

Analogously, the entropy 
\begin{equation}
S =\chi \sum_{\Omega }\ln \left( 1+\chi \exp \left[ -\beta \left(
E_{\Omega }+u^{\prime }\left( \bar{n}\right) -\mu \right) \right] \right)+\frac{1}{T}\sum_{\Omega }\bar{n}_{\Omega }\left( E_{\Omega }+u^{\prime
}\left( \bar{n}\right) -\mu \right),  \label{eq:Entropy1}
\end{equation}
and the mean energy reads
\begin{equation}
\bar{E}=\sum_{\Omega }\bar{n}_{\Omega }E_{\Omega }+U\left( V,\bar{n}\right) .
\label{eq:Energy1}
\end{equation}
Note that one could naturally expect above result. This occurs because the internal energy is the average of the kinetic term plus the interactions energy. Moreover, it is worth mentioning that such thermal functions were recently calculated for Lorentz-violating systems \cite{tt0,tt1,tt2,tt3,tt4,tt6,aa2022particles} and others \cite{ie1,ie2,ie3,ie4}.

%%%%%%%%%%%%%%%%%%%%%%%%%%%%%%%%%%%%%%%%%%%%%%%%%%%%%%%%%%%%%%%%%%%%%%%%%%%%%%%%%%%%%%%%%%%%%%%%%%%%%%%%%%%%%%%%%%%%%%%%%%%%%%%%%%%%%%%%%%%%%%%%%%%%%%%%%%%%%%%%%%%%%%%%%%%%%%%%%%%%%%%%%%%%%%%%%%%%%%%%%%%%%%%%%%%%%%%%%%%%%%%%%%%%%%%%%%%%%%%%%%%%%%%%%%%%%%%%%%%%%%%%%%%%%%%%%%%%%%%%%%%%%%%%%%%%%%%%%%%%%%%%%%%%%%%%%%%%%%%%%%%%%%%%%%%%%%%%%%%%%%%%%%%%%%%%%%%%%%%%%%%%%%%%%%%%%%%%%%%%%%%%%%%%%%%%%%%%%%%

\subsection{Analytical results for angular constraints}\label{Sec:Ex2}

A feasible remark to our model is the application to angular constraints. We provide this analysis with the help of the \textit{Euler-MacLaurin} formula. In particular, we study an \textit{interacting} quantum gas constrained in an one-dimensional ring with radius $R$. The mean energy in this angular approach has to be determined through specific boundary conditions (periodic boundary conditions), which yield:%
\begin{equation}
E_{\eta }=\frac{\hbar ^{2}}{2\mathrm{M}R^{2}}\eta ^{2}.
\end{equation}%
The grand canonical potential is written as%
\begin{align}
\Phi & =-k_{B}T\chi \sum_{\eta =-\infty }^{\infty }\ln \left\{ 1+\chi \mathfrak{z}%
\exp \left[ -\frac{\beta \hbar ^{2}}{2\mathrm{M}R^{2}}\eta ^{2}\right] \right\}
+U\left( V,\bar{n}\right) -u^{\prime }\left( \bar{n}\right) \bar{N},%
\displaybreak[0]  \\
& =-2k_{B}T\chi \sum_{\eta =0}^{\infty }\ln \left\{ 1+\chi \mathfrak{z}\exp \left[
-\frac{\beta \hbar ^{2}}{2\mathrm{M}R^{2}}\eta ^{2}\right] \right\} -k_{B}T\chi \ln
\left\{ 1+\chi \mathfrak{z}\right\} +U\left( V,\bar{n}\right) -u^{\prime
}\left( \bar{n}\right) \bar{N}.\nonumber
\end{align}%
After some algebraic manipulations, above expression is given by
\begin{equation}
\Phi =\frac{\mathcal{L}}{\lambda }h_{\frac{3}{2}}\left( \mathfrak{z}\right)
-h_{1}\left( \mathfrak{z}\right) +U\left( V,\bar{n}\right) -u^{\prime
}\left( \bar{n}\right) \bar{N}.
\end{equation}%
where $\mathcal{L}=2\pi R$ and $\mathfrak{z}=ze^{-\beta u^{\prime }\left( \bar{n}%
\right) }$ which encapsulates the details of interaction. We can also notice that there is no boundary effect here. Remarkably, this particular case can be used to investigate the respective thermodynamic properties of conducting rings and other analogous systems. More so, if one does not consider interactions, the results presented in this section will reproduce those ones exhibited in Ref. \cite{Dai2004}. 

\subsection{Analytical results for the torus}

As we did before, we can get another \textit{analytical} solution. In this case, we focus on the torus knot. Thereby, the respective grand canonical potential is%
\begin{align}
\Phi & =-T\chi \sum_{\delta =-\infty }^{\infty }\ln \left\{ 1+\chi \mathfrak{z}%
\exp \left[ -\frac{\beta \hbar ^{2}F\left( \alpha ,\eta \right) }{%
2\mathrm{M}\mathfrak{a}^{2}p^{2}} \delta^{2}\right] \right\} +U\left( V,\bar{n}\right) -u^{\prime
}\left( \bar{n}\right) \bar{N},\displaybreak[0]  \notag \\ 
& =-2T\chi \sum_{\delta =1}^{\infty }\ln \left\{ 1+\chi \mathfrak{z}\exp \left[
-\frac{\beta \hbar ^{2}F\left( \alpha ,\eta \right) }{2\mathrm{M}\mathfrak{a}^{2}p^{2}}\delta ^{2}%
\right] \right\} -T\chi \ln \left\{ 1+\chi \mathfrak{z}\right\} +U\left( V,%
\bar{n}\right) -u^{\prime }\left( \bar{n}\right) \bar{N}.
\end{align}%
Or solving the sum after some manipulations, we obtain%
\begin{equation}
\Phi =\frac{ p}{\sqrt{F\left( \alpha ,\eta \right)} }\frac{\mathcal{L}}{\lambda }h_{\frac{3}{2}}\left( \mathfrak{z}\right)
-h_{1}\left( \mathfrak{z}\right) +U\left( V,\bar{n}\right) -u^{\prime
}\left( \bar{n}\right) \bar{N}. \label{finaleq}
\end{equation}%
where $\mathcal{L}=2\pi\mathfrak{a}$ and again
the interaction contribution is also present in $h_{\sigma }\left( \mathfrak{z}%
\right) $ function. We can also notice that here there is no boundary effects for the same reason encountered in the ring case. Notice that Eq.(\ref{finaleq}) is modified by the topological function $F\left( \alpha ,\eta \right)$. On the other hand, the mean energy has the form
\begin{equation}
\mathcal{U}=\chi \frac{\mu \mathfrak{z}}{T}\log \left( 1+\chi e^{\mathfrak{z}}\right) +\frac{\lambda \mu \mathfrak{z}}{T%
}\frac{p\mathcal{L}}{\sqrt{F\left( \alpha ,\eta \right) }}h_{\frac{1%
}{2}}\left( \mathfrak{z}\right) +\frac{3\lambda }{2}\frac{p\mathcal{L}}{\sqrt{F\left( \alpha ,\eta \right) }}h_{\frac{3}{2}%
}\left( \mathfrak{z}\right)-h_{1}\left( \mathfrak{z}\right)+U\left( V,\bar{n}\right) -u^{\prime
}\left( \bar{n}\right) \bar{N}.
\label{eq:energyApp10}
\end{equation}

\begin{figure}[t]
\centering
\subfloat[Internal Energy]{
  \includegraphics[width=8cm,height=5cm]{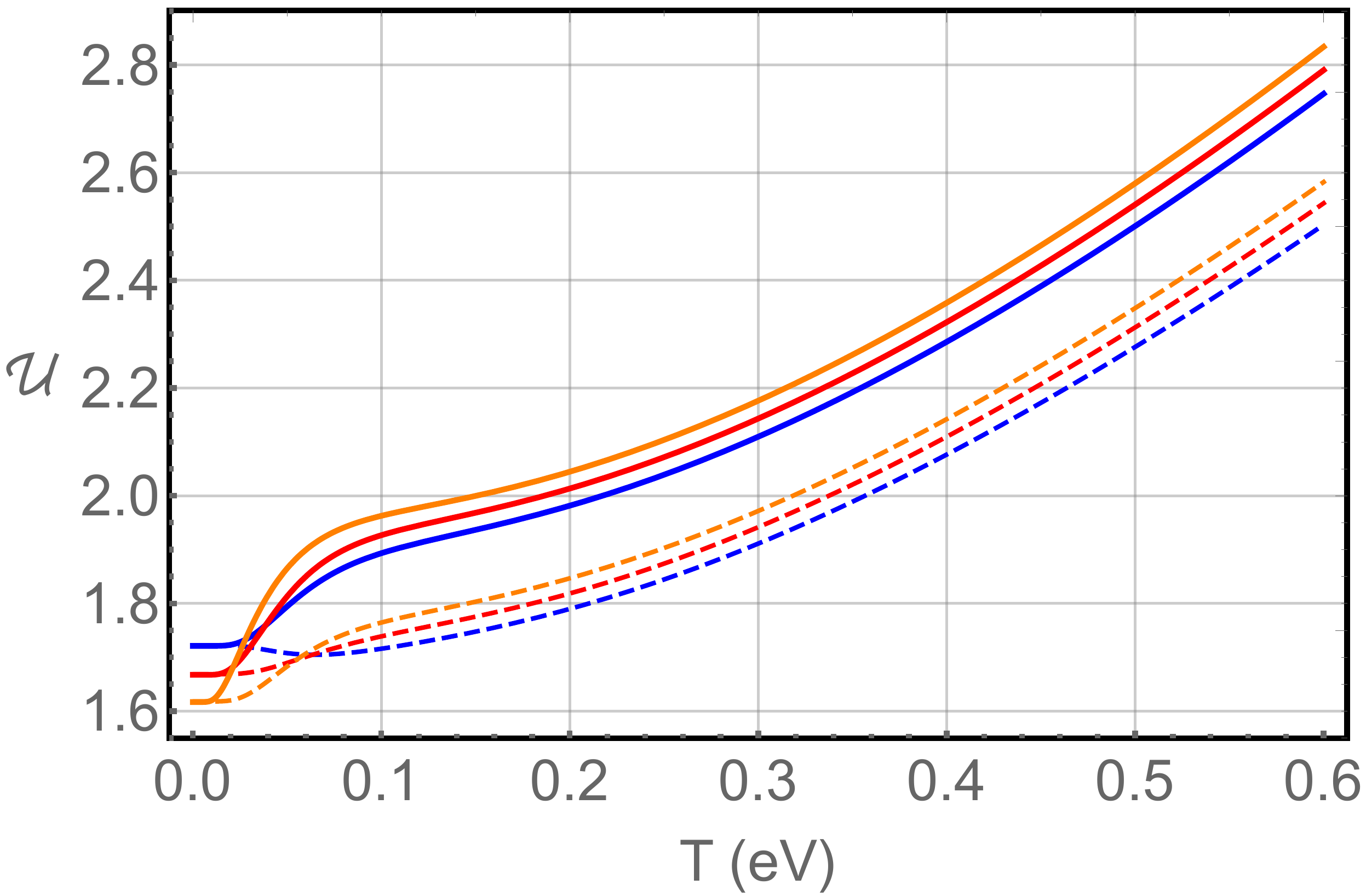}
  \label{fig:UApp-1}}
\subfloat[Internal Energy]{
  \includegraphics[width=8cm,height=5cm]{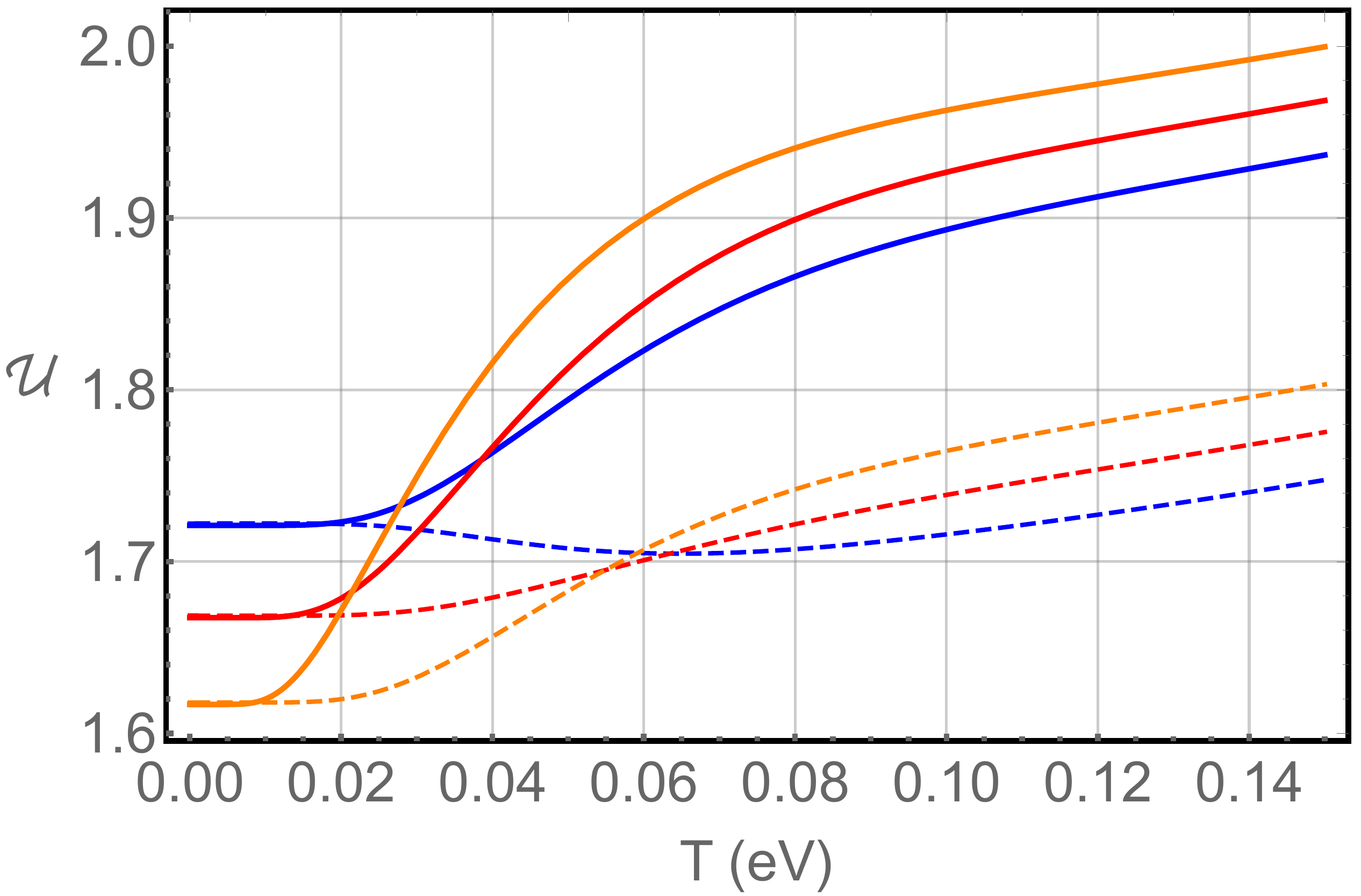}
  \label{fig:UApp-2}}
\caption{The internal energy for a linear approximation. The thick lines represents the system without interaction and the dashed line the interacted system in the context of linear approximation. The blue lines represents $\alpha=1$, the red lines $\alpha=5$ and the orange lines $\alpha=7$. In Fig. \ref{fig:UApp-2}, we just give a zoom to a better visualizations of the inversion points.}
\label{fig:UApp}
\end{figure}

In order to have a better comprehension of above equation, we provide a plot for fermions in the well-known linear approximation. This means that we keep up to firth order terms on the interaction function $u$ and we approximate the density to the well-known density of an ideal Fermi gas:
\begin{equation}
    n=\frac{1}{\sqrt{2}}\left(\frac{mT}{\pi} \right)^{\frac{3}{2}} h_{\frac{3}{2}}(z).
\end{equation}
Proceeding as mentioned, we find the plots displayed in Fig. \ref{fig:UApp}. The plots compare the situation with and without the interaction for three different values of the winding numbers. We first realize that the energies in the context of interactions have their values decreased when compared with the non-interacted one. This is caused mainly by the last term present in Eq. (\ref{eq:energyApp10}) which contributes with a negative term. We also see that all inversion points were shifted for a higher and approximated value of temperature $T\approx 0.06$ $\mathrm{eV}$.   

\begin{figure}[tbh]
\centering
\subfloat[Heat capacity]{
  \includegraphics[width=8cm,height=5cm]{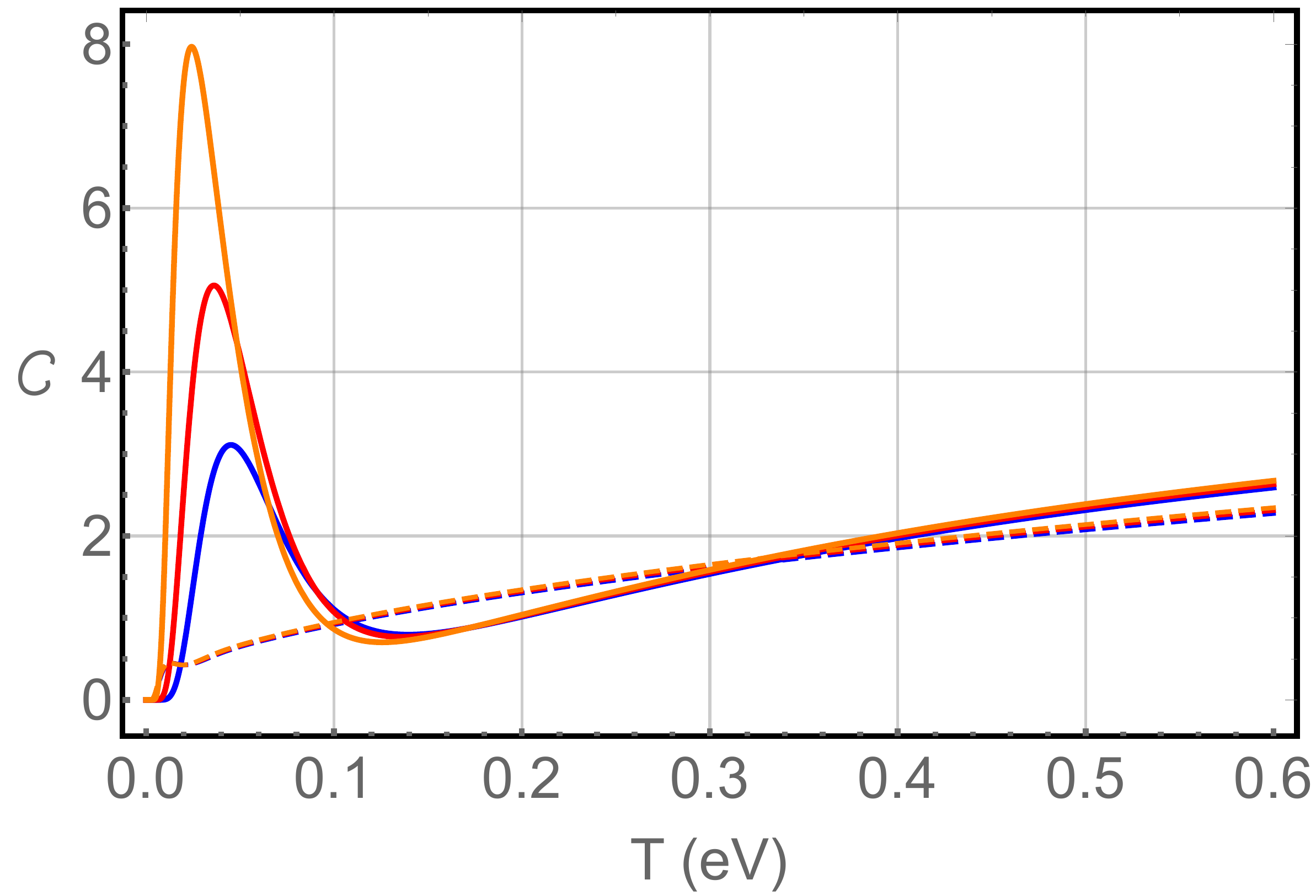}
  \label{fig:CApp-1}}
\subfloat[Heat capacity]{
  \includegraphics[width=8cm,height=5cm]{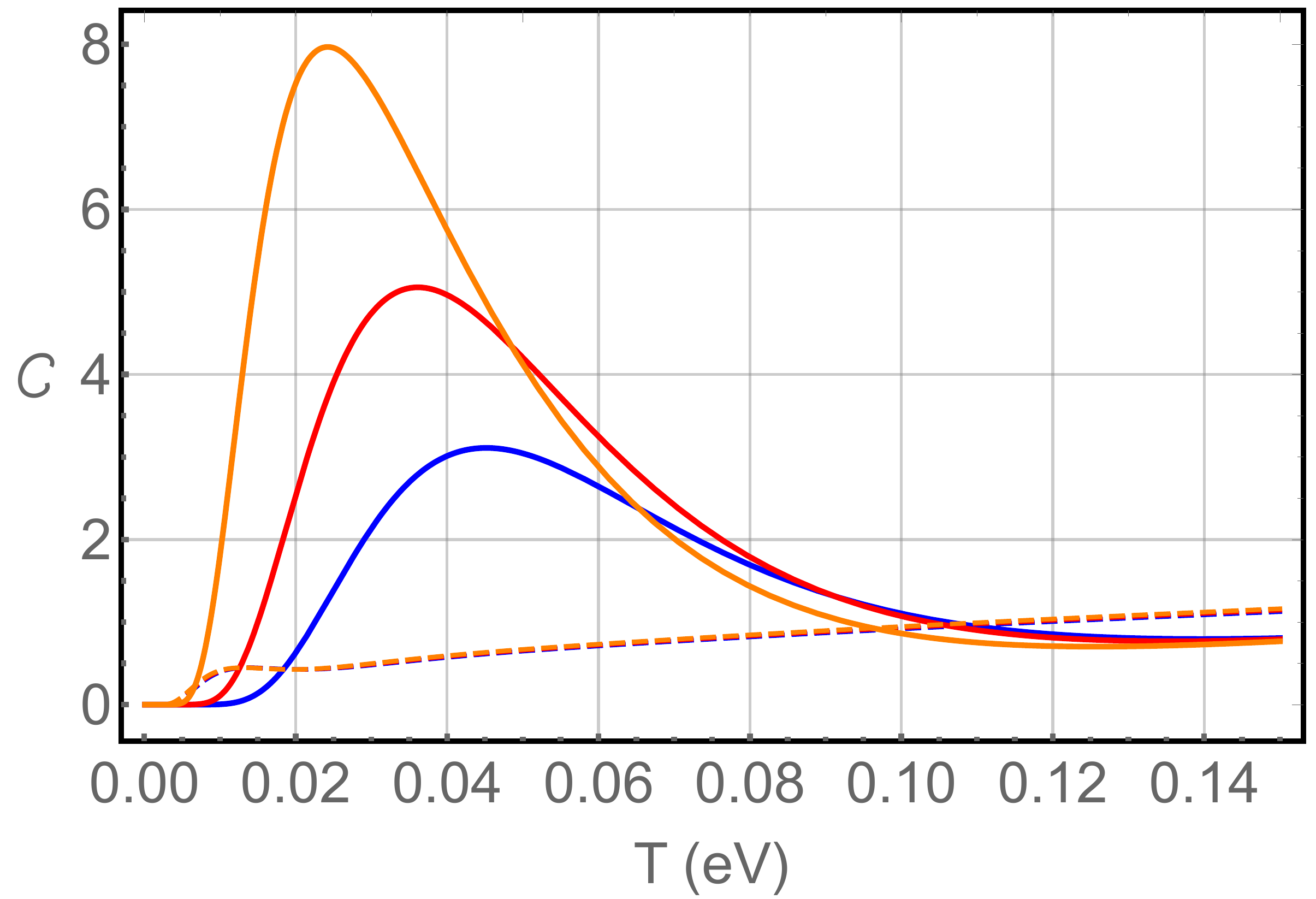}
  \label{fig:CApp-2}}
\caption{The heat capacity for a linear approximation. The thick lines represents the system without interaction and the dashed line the interacted system in the context of linear approximation. The blue lines represents $\alpha=1$, the red lines $\alpha=5$ and the orange lines $\alpha=7$. In Fig. \ref{fig:CApp-2} we restrict the range of temperature to a better visualizations of the behavior of the heat capacity for very low $T$. }
\label{fig:CApp}
\end{figure}

In possession of the energy, we can also calculate the heat capacity. The analytical result is omitted here because it is lengthy and it does not add anything useful to our discussion. However, the plots can bring some light to our understanding. In Fig. \ref{fig:CApp}, we display the plots for the heat capacity for different values of winding numbers. We can realize that the mound that once appeared in the situation without interaction fades as the interaction term is taken into account.

Finally, the number of particles is given by
\begin{equation}
N=\frac{\chi \mathfrak{z}}{T}\frac{p\mathcal{L}}{\sqrt{F\left( \alpha ,\eta \right)} }\log
\left( 1+\chi e^{\mathfrak{z}}\right) +\frac{\lambda \mathfrak{z}}{T}\frac{p\mathcal{L}}{\sqrt{F\left(
\alpha ,\eta \right) }}h_{\frac{3}{2}}\left( \mathfrak{z}\right).
\end{equation}%
This equation can be useful to calculate, for instance, the Fermi energy level of the system. The Fermi energy cannot be calculated from this equation until we conveniently choose the interaction $u(n)$ term. Nevertheless, we can at least see how the interaction and the topological parameters modify the
structure of the equation which determine the Fermi energy $\mu _{0}$. Even for the linear approximation, it is not possible to get an analytical results for this energy.

\begin{figure}[tbh]
\centering
  \includegraphics[width=8cm,height=5cm]{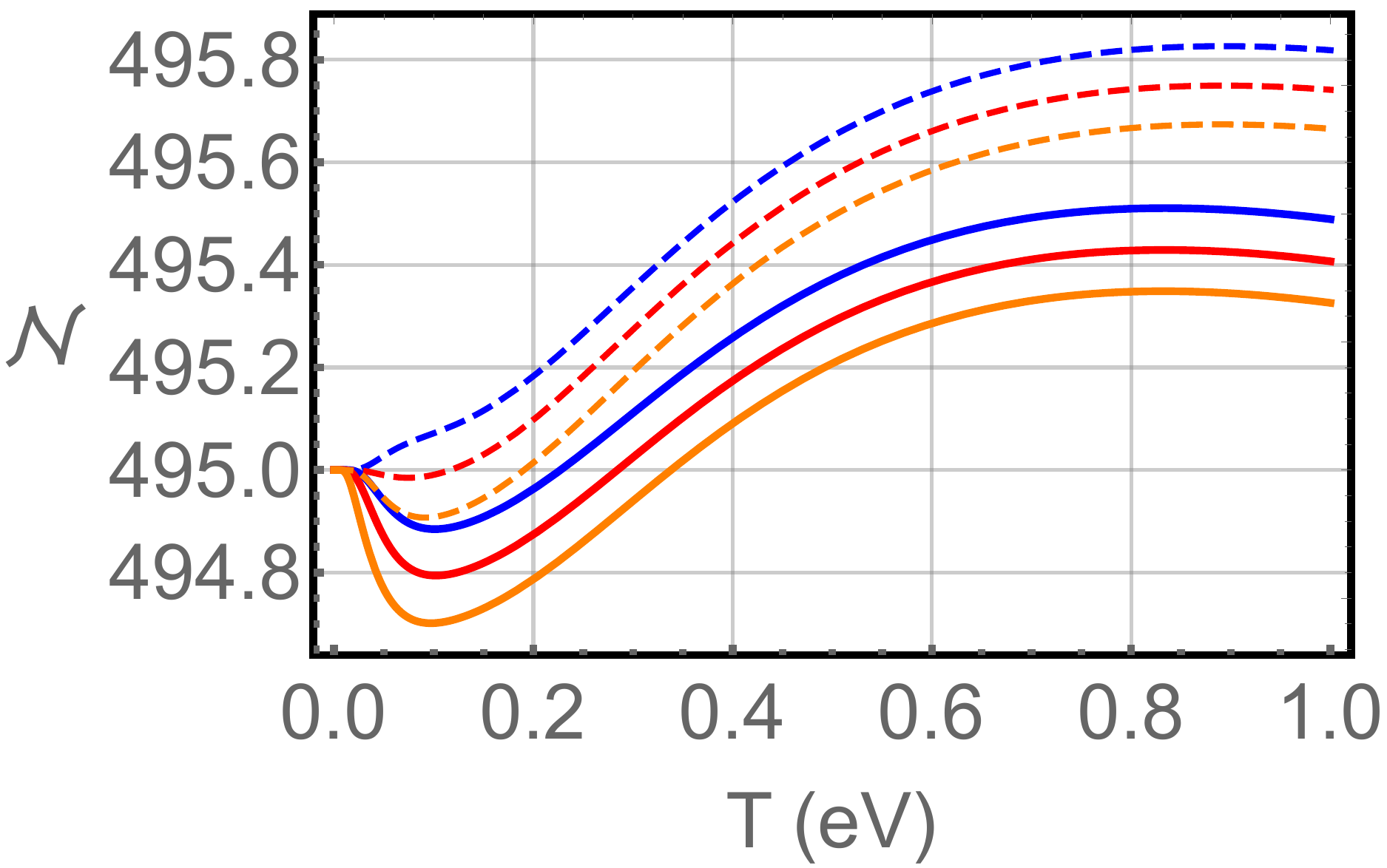}
\caption{The mean particle number in the low-temperature regime for different values of winding number $\alpha$ for the torus knot. The blue lines represents $\alpha=1$, the red lines $\alpha=5$ and the orange lines $\alpha=7$. Also the thick lines represents the system without interaction and the dashed line the interacted system in the context of linear approximation.}
\label{fig:Torus-C-LowT}
\end{figure}

However, we can plot the result numerically in order to get some insights. In Fig. \ref{fig:Torus-C-LowT}, we display the number of particle as a function of temperature for both scenarios, with (thick lines) and without (dashed lines) interactions for different winding numbers. At this point, we must remember that in the grand canonical approach the number of particle can fluctuate because the system is in contact with a thermal reservoir. We realise that the plots displayed, in Fig. \ref{fig:Torus-C-LowT}, the particle number varies with both temperature and winding number. As we can see, for a fixed temperature, the particle number increases as the winding number decreases. Also, we observe that, in the range $0.01~\mathrm{eV}<T<0.1~\mathrm{eV}$, the number of particle decreases until achieve its minimum value around $T=0.1~\mathrm{eV}$. After, it starts to increase. When we are dealing with interactions, most of the behavior is maintained; however, we realise that the fluctuations have their values increased when compared to the non-interacting scenario.

%%%%%%%%%%%%%%%%%%%%%%%%%%%%%%%%%%%%%%%%%%%%%%%%%%%%%%%%%%%%%%%%%%%%%%%%%%%%%%%%%%%%%%%%%%
\pagebreak

%%%%%%%%%%%%%%%%%%%%%%%%%%%%%%%%%%%%%%%%%%%%%%%%%%%%%%%%%%%%%%%%%%%%%%%%%%%%%%%%%%%%%%%%%%

\section{Conclusion and future perspectives}\label{conclusion}

We examined the behavior of the thermodynamic functions for a torus knot. Initially, non interacting gases were taken into account with the usage of the grand canonical ensemble description. A study of how this geometry affects the system of fermionic and bosonic particles was provided as well.

We observed that for the bosonic sector the internal energy increased with both temperature and winding number. Also, it is worth mentioning that for fermions, there was a set of “inversion point” between the range $0.02~\mathrm{eV}<T<0.04~\mathrm{eV}$ due to the winding number. We observed that it shifted the inversion point to larger values of temperature. Then, we see a concrete modification in the thermal properties due to the topological parameter $\alpha$.

We also saw modifications due to the winding number in the heat capacity. For bosons, we have obtained that it increased with $\alpha$. For fermions, on the other hand, we had the formation of a mound between the range $0.01~\mathrm{eV}<T<0.1~\mathrm{eV}$. At this range, the heat capacity reached its maximum value.  

We also constructed a model to provide a description of interacting quantum gases; it was implemented to a ring and to a torus. Such an interaction sector turned out to be more robust since all results were derived analytically. We derived analytical expressions for the internal energy and for the particle number of a torus. With these results, we also explored the well known linear approximation. In this context, we compared the internal energy for the system with and without interactions. In particular, when interactions were turned on, the energy had been changed down and the inversion point for different values of $\alpha's$ was all shifted to $T\approx0.06~\mathrm{eV}$. For the heat capacity, the topological parameter almost played no role in our study. 

Finally, we displayed some plots for the number of particles and compared the situations with and without the linear approximation. We obtained that the particle number had a minimum at $T=0.1~\mathrm{eV}$. We also observed that the values increased when the interactions were turned on. The winding number also changed the fluctuations of the number of particle. We see that as the winding number increased, the particle number decreased. This behavior remained in the scenario with interaction. Furthermore, another remarkable feature worth exploring, would be the thermodynamic aspects of anisotropic systems \cite{kostelecky}.

%%%%%%%%%%%%%%%%%%%%%%%%%%%%%%%%%%%%%%%%%%%%%%%%%%%%%%%%%%%%%%%%%%%%%%%%%%%%%%%%%%%%%%%%%

%%%%%%%%%%%%%%%%%%%%%%%%%%%%%%%%%%%%%%%%%%%%%%%%%%%%%%%%%%%%%%%%%%%%%%%%%%%%%%%%%%%%%%%%%%
\section*{Acknowledgments}
\hspace{0.5cm}

The authors also express their gratitude to FAPEMA, CNPq and CAPES (Brazilian research agencies) for invaluable financial support. In particular, J.A.A.S. Reis is supported by FAPEMA BPD-08734/22, and A. A. Araújo Filho is supported by Conselho Nacional de Desenvolvimento Cientíıfico e Tecnológico (CNPq) -- 200486/2022-5. Most of the calculations were performed using the \textit{Mathematica} software. 

%%%%%%%%%%%%%%%%%%%%%%%%%%%%%%%%%%%%%%%%%%%%%%%%%%%%%%%%%%%%%%%%%%%%%%%

%%%%%%%%%%%%%%%%%%%%%%%%%%%%%%%%%%%%%%%%%%%%%%%%%%%%%%%%%%%%%%%%%%%%%%%%%%%%%%%%

%-------------------------------------------------------------------------

\bibliographystyle{ieeetr}
\bibliography{ref}

\end{document}